\documentstyle[epsfig,aps,preprint]{revtex}
\begin{document}
\tightenlines
\draft
\title{Analytical results for string propagation near a Kaluza-Klein
black hole}
\author{H.~K.~Jassal\thanks{E--mail : hkj@ducos.ernet.in} and
A.~Mukherjee\thanks{E--mail : am@ducos.ernet.in}}
\address{Department of Physics and Astrophysics, \\
	 University of Delhi, Delhi-110 007, India.
	}
\maketitle
\begin{abstract}
This brief report presents analytical solutions to
the equations of motion of a null string.
The background spacetime is a magnetically charged Kaluza-Klein black
hole. 
The string coordinates are expanded with the world-sheet velocity of
light as an expansion parameter.
It is shown that the zeroth order solutions can be expressed in terms
of elementary functions in an appropriate large distance
approximation.  
In addition, a class of exact solutions corresponding to the
Pollard-Gross-Perry-Sorkin monopole case is also obtained. 
\end{abstract}
\
String propagation in curved spacetime has been a subject of study in
a large number of papers (For a review, see \cite{erice}). 
To study string interaction with the background spacetime,
the string equations of motion are obtained from the world-sheet action
and are solved by simplifying them with suitable ansatze. 
It has been  suggested by de Vega and Nicolaidis \cite{vega} that 
the world-sheet velocity of light $c$, the velocity of wave
propagation along the string, can be an expansion parameter for the
string coordinates. 
The limit of small $c$ corresponds to the strong gravity limit
\cite{vega,vega2,sanchez}, the length of the string being much larger
than the curvature radius. 

The formalism developed in the literature
\cite{erice,vega,vega2,sanchez} is applicable to both cosmic strings
and fundamental strings. 
However, in the latter case the theory is formulated in higher
dimensions.
Thus we expect a nontrivial contribution to the string dynamics from
the extra compact dimensions.
In other words, it is of interest to study in what way the propagation
of a string probe is affected by these extra dimensions.
An attempt in this direction was made in \cite{kkbh}.
The propagation of a null string in a Kaluza-Klein black
hole background  was studied.
It was shown that, even at the classical level, the unfolding of an extra
dimension can be seen by the string probe.
The Kaluza-Klein radius, i.e., the radius around which the extra dimension
winds, decreases for a magnetically charged black hole, as viewed by a null
string approaching it.
For an electrically charged background, the Kaluza-Klein radius behaves
in the opposite sense.

In this Brief Report we extend the work done already in
ref. \cite{kkbh} and obtain analytical counterparts of the solutions
presented there for the magnetically charged black hole case, for
small values of the scalar charge. 
In addition, we show that the problem is solvable exactly for the case
of the Pollard-Gross-Perry-Sorkin (PGPS) monopole,
which has been shown recently to arise as  a solution of a suitable
dimensionally reduced string theory \cite{sroy}. 
This report is more or less technical in content. 
We believe our results may be of some significance since analytical
solutions are presented for physically interesting higher dimensional
backgrounds.  

We start with the  string world-sheet action \cite{green} given by
\begin{equation}
S = -T_{0}\int d\tau d \sigma \sqrt{-detg_{ab}}
\end{equation}
where $g_{ab}=G_{\mu \nu}(X)\partial_{a}X^{\mu}\partial_{b}X^{\nu}$ is
the two-dimensional world-sheet metric; $\sigma$ and $\tau$ are the
world sheet coordinates and $T_0$ is the string tension.

The classical equations of motion in the conformal
gauge \cite{vega} are given by 
\begin{equation}
\partial_{\tau}^{2}X^{\mu}-c^{2}\partial_{\sigma}^{2}X^{\mu}+\Gamma_{\nu
\rho}^{\mu} \left[\partial_{\tau}X^{\nu}\partial_{\tau}X^{\rho} -
c^{2}\partial_{\sigma}X^{\nu}\partial_{\sigma}X^{\rho}\right]=0 .
\end{equation}
and the constraints are given by
\begin{equation}
\partial_{\tau}X^{\mu}\partial_{\sigma}X^{\nu}G_{\mu \nu} = 0
\end{equation} 
\begin{equation}
[\partial_{\tau}X^{\mu}\partial_{\tau}X^{\nu} + c^{2}\partial_{\sigma}X^{\mu}
\partial_{\sigma}X^{\nu}]G_{\mu \nu} = 0,
\end{equation}
\noindent $c$ being the velocity of wave propagation along the string.

The string coordinates are expanded perturbatively with the world-sheet
velocity of light as the expansion parameter, {\it viz.},
\begin{equation}
X^{\mu}(\sigma, \tau) = X_{0}^{\mu}(\sigma, \tau) + c^{2} X_{1}^{\mu}(\sigma,
\tau) + c^{4} X_{2}^{\mu}(\sigma, \tau) + ..., 
\end{equation}
The zeroth order coordinates $X_{0}^{\mu}(\sigma,\tau)$ satisfy the following
set of equations
\begin{eqnarray}
\ddot X_{0}^{\mu} + \Gamma_{\nu \rho}^{\mu} \dot X_{0}^{\nu} \dot X_{0}^{\rho} & =
& 0, \\ \nonumber
\dot X_{0}^{\mu} \dot X_{0}^{\nu} G_{\mu \nu} & = & 0, \\ \nonumber
\dot X_{0}^{\mu} X_{0}'^{\nu} G_{\mu \nu} & = & 0,
\end{eqnarray}
where dot and prime denote differentiation w.r.t. $\tau$ and $\sigma$
respectively. 
These equations describe the motion of a null string \cite{vega}. 
The second equation constrains the motion to be perpendicular to the
string. 

We study string propagation in a background given by the metric
\cite{gibbons} 
\begin{equation}
ds^{2}=-e^{4 k \frac{\varphi}{\sqrt{3}}} (dx_{5} +
2 k A_{\alpha}dx^{\alpha})^{2} + e^{-2 k \frac{\varphi}{\sqrt{3}}}g_{\alpha
\beta}dx^{\alpha}dx^{\beta},
\end{equation}
where $k^{2}=4 \pi G$; $x_{5}$ is the extra dimension and winds around
a circle, $\varphi$ and $A_{\mu}$ represent the scalar field and the
gauge field respectively.
Here $g_{\alpha \beta}$ is the four-dimensional metric.

The mass $M$ of the black hole, the electric charge $Q$, the
magnetic charge $P$ and the scalar charge $\Sigma$ are constrained by
\begin{equation}
\frac{2}{3} \Sigma = \frac{Q^{2}}{\Sigma + \sqrt{3} M} +
\frac{P^{2}}{\Sigma - \sqrt{3} M}, \label{constr}
\end{equation}
where the scalar charge is defined by  \\
\begin{eqnarray}
k\varphi \longrightarrow \frac{\Sigma}{r} +
O\left(\frac{1}{r^2}\right) \mathrm{as}~~~~ r \longrightarrow \infty. \nonumber
\end{eqnarray}

The black hole solutions, in the notation of ref. \cite{itzhaki}, are
\begin{eqnarray}
e^{4 \varphi/\sqrt{3}} = \frac{B}{A}, A_{\mu}dx^{\mu} &=&
\frac{Q}{B}(r-\Sigma)dt + P \cos\theta d\phi \\ \nonumber
g_{\mu \nu}dx^{\mu}dx^{\nu} = \frac{f^2}{\sqrt{AB}} dt^{2} &-&
\frac{\sqrt{AB}}{f^2} dr^{2}  \\ \nonumber 
&-& \sqrt{AB} \left(d\theta^2 + \sin^2\theta
d\phi^2\right) ,
\end{eqnarray}
with $A$, $B$ and $f$ given by
\begin{eqnarray}
A & = & (r-\frac{\Sigma}{\sqrt{3}})^{2} - \frac{2
P^{2}\Sigma}{\Sigma-\sqrt{3} M}  \\ \nonumber
B & = & (r+\frac{\Sigma}{\sqrt{3}})^{2} - \frac{2 Q^{2} \Sigma}{\Sigma
+ \sqrt{3} M}\\ \nonumber
f^{2} & = & (r-M)^{2} - (M^{2} + \Sigma^{2} - P^{2} - Q^{2}) 
\end{eqnarray}

The zeroth order equations of motion for the string coordinates in
a purely  magnetically charged background ($Q=0$) are 
\begin{eqnarray}
\frac{\partial^{2}t}{\partial \tau^{2}} &+& 2\left(\frac{f'}{f} -
\frac{B'}{2B}\right) \frac{\partial t}{\partial \tau} \frac{\partial
r}{\partial \tau} =0, \\ \nonumber
\frac{\partial^{2}r}{\partial \tau^{2}} &+& \left[
-\frac{f^{3}}{2AB^{2}}(B'f-2f'B) \right] \left( \frac{\partial
t}{\partial \tau} \right)^{2} \\ \nonumber
&+& 
\left(\frac{A'f -  2f'A}{2Af} \right)
\left( \frac{\partial r}{\partial \tau} \right)^{2}
- 
\frac{f^{2}A'}{2A} \left( \frac{\partial \phi}{\partial \tau}
\right)^{2} \\ \nonumber
&+& 
\frac{f^2}{2A^{3}}(A'B-B'A) \left( \frac{\partial
x_{5}}{\partial \tau} \right)^{2} =0, \\ \nonumber
\frac{\partial^{2}\phi}{\partial \tau^{2}} &+& \frac{A'}{A}
\left(\frac{\partial r}{\partial \tau} \right) \left(\frac{\partial
\phi}{\partial \tau} \right) = 0, \\ \nonumber
\frac{\partial^{2} x_{5}}{\partial \tau^{2}} &+& \left(-\frac{A'}{A}
+ \frac{B'}{B} \right) \left(\frac{\partial r}{\partial
\tau}\right)\left(\frac{\partial x_{5}}{\partial \tau}\right) = 0.  \label{eoms}
\end{eqnarray}
Here we restrict ourselves to the exterior region $r>M$ and the
equatorial plane, i.e. $\theta=\pi/2$. 

These equations can be reduced to quadratures involving 
constants of integration which depend on $\sigma$:   
\begin{eqnarray}
\tau &=& \int \frac{dr}{\sqrt{\frac{B}{A}c_{1}^2 - \frac{f^2}{B}
c_{3}^2}}, \\ \nonumber 
x_{5} &=& \int \frac{c_{3}A dr}{B\sqrt{\frac{B}{A}c_{1}^2 -
\frac{f^2}{B} c_{3}^2}}, \\ \nonumber 
t &=& \int \frac{ c_{1} B dr}{f^2 \sqrt{\frac{B}{A}c_{1}^2 -
\frac{f^2}{B} c_{3}^2}}, \label{integs}
\end{eqnarray}
and can be solved to obtain $t$, $r$, $\phi$ and $x_{5}$ as functions
of $\tau$ \cite{kkbh}.  
In (12) we have taken the case of a string falling in `head-on'.

The integrals (12) have been evaluated numerically
\cite{kkbh} and inverted to obtain the coordinates as functions of
$\tau$ in the limit $r >>\Sigma$.
However, the quadratures can be reduced to combinations of elliptical
integrals, depending on the relative values of constants $\Sigma_1$
and $M$ (see \cite{integtab}).
Rather than tabulate these, we concentrate on the region $r>>\Sigma$,
where the solutions reduce to elementary functions.
We take for instance the case $c_1=c_3=1$. 
Up to the first order in $\Sigma_1/r$, the solutions (modulo integration
constants which depend on $\sigma$) are 
\begin{eqnarray}
\tau &=& -\frac{2}{3 \sqrt{2(M+\Sigma_1)}} \left\{r + \frac{M
\Sigma_1}{M + 3 \Sigma_1} \right\}^{3/2} \\ \nonumber
x_5 &=& -\frac{2}{\sqrt{2(M+3\Sigma_1)}}
\left[\left(\frac{r}{3}+\alpha-\frac{9\Sigma_1}{2}\right)(r-\Sigma_1)^{1/2}\right] \\ \nonumber
&-& \frac{\left(7\Sigma_1-\alpha \right)^{3/2}}{\sqrt{2(M+\Sigma_1)}}
\mathrm{tan}^{-1}\left\{\frac{\sqrt{2(r-3\Sigma_1)}}{\sqrt{7\Sigma_1-\alpha}}\right\}
\\ \nonumber
t &=&
-\frac{2}{\sqrt{2(M+3\Sigma_1)}}\left[(r-\alpha)\left\{\frac{r}{3}+2M+\Sigma_1\right\}
\right] \\ \nonumber
&+&\frac{2(2M+\Sigma_1)^2}{\sqrt{2(M+3\Sigma_1)}\sqrt{\alpha-2M-\Sigma_1}}
\times \\ \nonumber
& & \mathrm{tan}^{-1}\left\{\frac{\sqrt{r-\alpha}}{\sqrt{\alpha-2M-\Sigma_1}}\right\}
\end{eqnarray}
where $\Sigma_1=\Sigma/\sqrt{3}$ and
$\alpha=\frac{3M\Sigma_1}{3\Sigma_1+M}$.
These solutions are valid in the region outside the horizon but not
asymptotically far from the black hole.
The negative sign comes because we consider an in-falling string.
The solutions match with the numerical solutions reported in
\cite{kkbh} for the corresponding values of $c_1$ and $c_3$ (see
Figure 1).

Equations (11)  can be solved exactly, i.e., without resorting
to the limit $r>>\Sigma$, for the special case  $P=2M$ and $Q=0$ (the
Pollard-Gross-Perry-Sorkin monopole \cite{gross}). 
The constraint equation (\ref{constr}) implies that $\Sigma_1=-M$ or
$\Sigma_1=2M$. 
In the former case, the metric reduces to the form  reported in
ref.\cite{gross}. 
Here the time coordinate $t$ is the same as the proper time
$\tau$ of the string.  
The solutions are
\begin{eqnarray}
\tau=t &=& \frac{1}{\sqrt{1-c_{3}^{2}}}
(r-\beta M)^{1/2}(r+3M)^{1/2} \\ \nonumber
&+&  \frac{(3+\beta)M}{\sqrt{1-c_{3}^{2}}} \mathrm{ln}\left[(r-\beta
M)^{1/2}+(r+3M)^{1/2}\right]  \\ \nonumber
x_5 &=& \frac{1}{\sqrt{1-c_{3}^{2}}}
(r-\beta M)^{1/2}(r+3M)^{1/2} \\ \nonumber
&+& \frac{(11+\beta)M}{\sqrt{1-c_{3}^{2}}} \mathrm{ln}\left[(r-\beta
M)^{1/2}+(r+3M)^{1/2}\right]  \\ \nonumber
&+&\frac{16M}{\sqrt{\beta-1}\sqrt{1-c_{3}^{2}}}
\left[\mathrm{arctan}\frac{2\sqrt{r-\beta
M}}{\sqrt{\beta-1}\sqrt{r+3M}} \right]
\end{eqnarray}
where $\beta=\frac{1+3c_{3}^{2}}{1-c_{3}^{2}}$. 
We choose $c_1=1$; the condition of reality of the solutions then
forces $c_3<1$ and consequently $\beta>1$. 
Figure 2 shows a plot of $r$ versus $\tau$ for $c_3=0.6$.
A comparison of figures 1 and 2 shows that the approach to the horizon
is different in the two  cases. 
In the PGPS case, the string decelerates as it approaches the
horizon. 
This is not surprising as the `repulsive' or `anti-gravity` effect of
extremal black holes has been commented on in the literature (see for
example, \cite{thooft} and  \cite{gibb_BPS}). 

In addition to the above cases, there is another solution (which has
not been mentioned hitherto in the literature) corresponding to
$\Sigma_1=2M$. 
The integrals can be solved in terms of elliptical functions, the
solutions being  
\begin{eqnarray}
\tau &=& \frac{1}{3}\sqrt{\frac{2}{7M}} \sqrt{(r-6M)(r+2M)
\left(r-\frac{10M}{7}\right)} \\ \nonumber
&-&
\frac{32M}{21\sqrt{7}}\left[\left\{\mathrm{E}\!\left(g(r),\frac{3}{7}\right)
-4\mathrm{F}\!\left(g(r),\frac{3}{7}\right)\right\}\right] 
\\ \nonumber
x_5 &=&  \frac{1}{3}\sqrt{\frac{2}{7M}} \sqrt{(r-6M)(r+2M)
\left(r-\frac{10M}{7}\right)} \\ \nonumber
&-&
\frac{32M}{21\sqrt{7}}\left[\left\{22~\mathrm{E}\!\left(g(r),\frac{3}{7}\right)
-4\mathrm{F}\!\left(g(r),\frac{3}{7}\right)\right\} \right] \\ \nonumber 
t &=&  \frac{1}{3}\sqrt{\frac{2}{7M}}
\sqrt{(r-6M)(r+2M)\left(r-\frac{10M}{7}\right)} \\ \nonumber
&-&
\frac{2M}{21\sqrt{7}}\left[52~~\mathrm{F}\!\left(g(r),\frac{3}{7}\right)
\right] \\ \nonumber 
&+&\frac{M}{21\sqrt{7}}\left[59\left\{8M
~\mathrm{E}\!\left(g(r),\frac{3}{7}\right) 
- 6M
~\mathrm{F}\!\left(g(r),\frac{3}{7}\right)\right\}\right] \\ \nonumber
&+&\frac{2M}{21\sqrt{7}}\left[63~~\mathrm{\Pi}\!
\left(\frac{4}{7},g(r),\frac{3}{7}\right)\right],
\end{eqnarray}
where $g(r)=\mathrm{arcsin}\left[\frac{1}{2} \sqrt{\frac{7}{6}}
\sqrt{\frac{r+2M}{M}} \right]$. 
In these expressions $\mathrm{F}$, $\mathrm{E}$ and $\mathrm{\Pi}$ are
elliptical functions of first, second and third kind respectively(see
\cite{integtab}).
We have chosen $c_1=c_3=1$ in this case.
\vskip 0.5cm
We have found out analytical solutions to the equations of motion to
zeroth order in the expansion in powers of $c$.
The solutions obtained in the case when $r$ is large compared to the
scalar charge match with those obtained numerically in \cite{kkbh}, for
corresponding values of the integration constants $c_1$ and $c_3$.
Although the results presented  for the large $r$ case are
technically the same as the ones obtained in \cite{kkbh}, the
analytical solutions reveal two new cases where the problem is solved
exactly. 
These include the well known PGPS extremal black hole.

The equations of motion for the electrically charged black hole case do
not reduce to quadratures (see \cite{kkbh}).
However, the analytical results obtained in the magnetically charged
case give hope that some qualitative predictions can be made in
that case as well.
The optimism comes from the fact that the electrically charged black
hole solutions are related to the magnetically charged ones by the
duality transformation $P\rightarrow Q$, $Q\rightarrow P$ and
$\Sigma\rightarrow -\Sigma$ \cite{gibbons,itzhaki}. 
These analytical zeroth order results also form the basis for higher
order calculations, work on which is in progress.

\acknowledgements
The work presented here is a continuation of a collaboration with the
late Professor R.~P.~Saxena who passed away in June, 1999.
The authors acknowledge useful discussions with Parthasarathi Majumdar.
Thanks are also due to Shobhit Mahajan for help with symbolic
manipulation software. 
H.~K.~J. thanks the University Grants Commission, India, for a
fellowship.

\begin{figure}
\parbox{6.5cm}
{
\epsfig{file=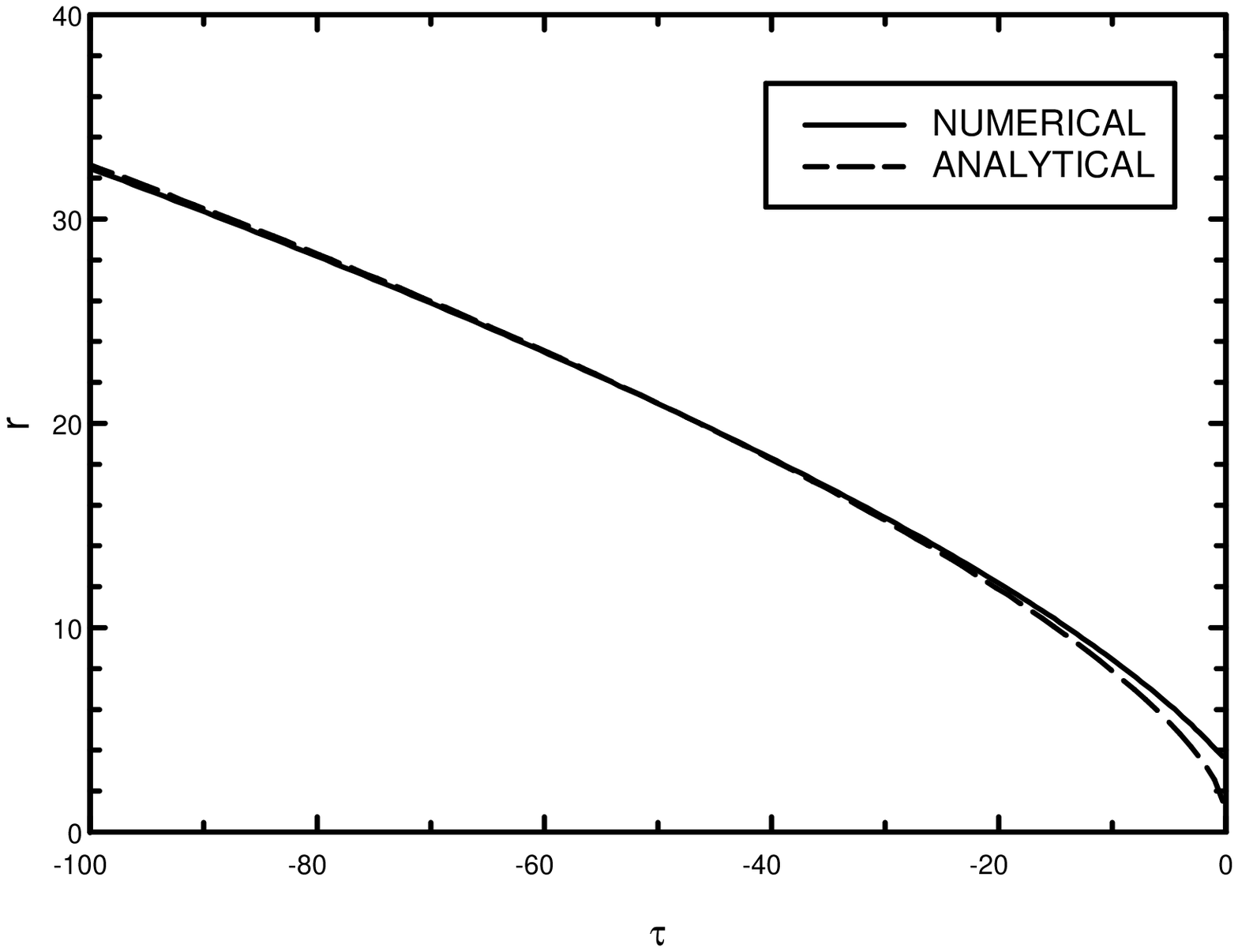,height=20.0cm,width=15.0cm}
\caption{Plot of $r$ versus $\tau$ showing analytical and numerical
results for $r>>\Sigma_1$ and $c_1=c_3=1$} 
}
\end{figure}
\begin{figure}
\parbox{6.5cm}
{
\epsfig{file=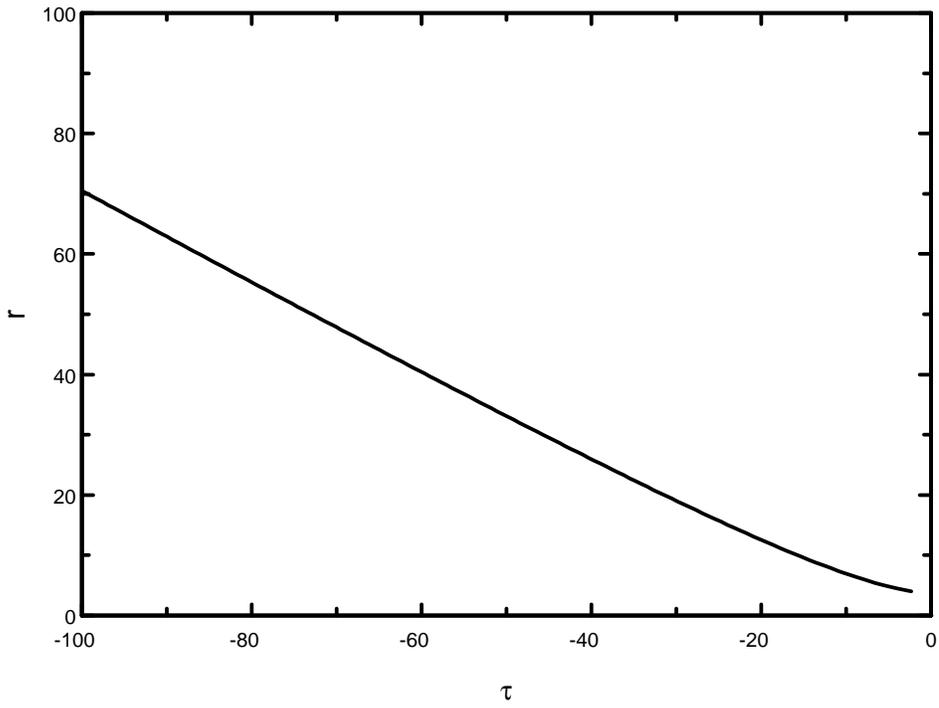,height=20.0cm,width=15.0cm}
\caption{$r$ versus $\tau$ when $\Sigma_1=-M$ for $c_1=1$ and $c_3=0.6$}
}
\end{figure}

\end{document}